\documentclass[12pt]{article}
\usepackage{a4,latexsym,graphicx,epsfig,psfrag,here}
\normalsize 
\begin{document}
\begin{titlepage} 
\begin{flushright} 
UWThPh-2001-29\\ 
August 2001\\ 
\end{flushright} 
\vspace{2.5cm} 
\begin{center} 
{\Large \bf CHIRAL PERTURBATION THEORY \\[10pt] AND KAON PHYSICS$^*$
  }
\\[40pt]
Gerhard Ecker       \\
\vspace*{.5cm} 
Inst. Theor. Physik, Univ. Wien\\ 
Boltzmanng. 5, A-1090 Wien, Austria \\[10pt]
\end{center} 

\vspace*{1.5cm} 
\begin{abstract}\noindent
A brief survey of applications of chiral perturbation theory to both 
semileptonic and nonleptonic kaon decays  is presented. Special
emphasis is given to recent theoretical advances related to
semileptonic decays, in particular pion pion scattering and $K_{e4}$
decays. The systematic approach for including isospin violation and
electromagnetic corrections in semileptonic kaon decays is discussed
for $K_{l3}$ decays.
\end{abstract}

\vfill
\begin{center}
To appear in the Proceedings of\\[5pt]
KAON 2001, Pisa, Italy, June 12 - 17, 2001 \\[5pt]
\end{center}

\vfill 
\noindent *~Work supported in part by TMR, EC-Contract  
No. ERBFMRX-CT980169 (EURODA$\Phi$NE).
 
\end{titlepage} 
\addtocounter{page}{1} 
\section{Survey}
\renewcommand{\arraystretch}{1.3}
Chiral perturbation theory (CHPT) sets in where the operator product
expansion (OPE) stops (Table \ref{tab:Fermi}). 
Hadronic matrix elements of quark operators are calculated
with an effective field theory directly in terms of hadron
fields. The theory shares the symmetries of the effective
theory of three light quarks, gluons, photon and leptons
derived from the Standard Model via the OPE
at a scale of the order of $m_c$. The step from $m_c$
down to $M_K$ entails an enormous loss of information that is encoded
in the coupling constants (LECs) of the effective chiral Lagrangian  
${\cal L}_{\rm eff}^{\rm SM}$ in Table \ref{tab:EFTSM}. All the
Lagrangians in Table \ref{tab:EFTSM} are used in present-day CHPT 
calculations of kaon decays.
 
\begin{table}[t]
  \centering
  \caption{\it From the Fermi scale to $M_K$.}
  \vskip 0.1 in
\begin{tabular}{ccc}
% & & \\
\hline \\
 $M_W$ & $SU(3)\times SU(2)\times U(1)$  &  
quarks, gluons, leptons \\
 & SM  & gauge bosons, Higgs \\
% & & \\
OPE & $\Downarrow$ & perturbative \\
% & & \\
$m_c$ & ${\cal L}_{\rm QCD}^{N_f=3}, {\cal L}_{\rm eff}^{\Delta S=1}, 
\dots$ & light quarks, gluons \\
 & & leptons, photon \\
% & & \\
symmetries & $\Downarrow$ & nonperturbative \\
% & & \\ 
 $M_K$  &  ${\cal L}_{\rm eff}^{\rm SM}$  &  hadrons \\
 & (Table \ref{tab:EFTSM}) & leptons, photon \\
 & & \\
\hline
\end{tabular}
\label{tab:Fermi}
\end{table} 

\begin{table}[ht]
  \centering
  \caption{ \it Effective chiral Lagrangian  ${\cal L}_{\rm eff}^{\rm SM}$
 of the SM relevant for kaon physics. The numbers in brackets denote
  the numbers of low-energy constants.}
  \vskip 0.1 in
$$
\begin{tabular}{|l|c|} 
\hline
%&  \\
\hspace{1cm} ${\cal L}_{\rm chiral\;  order}$ 
~($\#$ of  LECs)  &  loop   ~order \\[3pt] 
\hline 
%&  \\[3pt]
${\cal L}_{p^2}(2)$~+~${\cal L}_{p^4}^{\rm odd}(0)$ & \\[3pt]
~+~${\cal L}_{G_Fp^2}^{\Delta S=1}(2)$  & $L=0$\\[3pt]
~+~${\cal L}_{e^2p^0}^{\rm em}(1)$~+~${\cal L}_{G_8e^2p^0}^{\rm
   emweak}(1)$ & \\[10pt]
~+~${\cal L}_{p^4}^{\rm even}(10)$~+
~${\cal L}_{p^6}^{\rm odd}(32)$ & \\[3pt]
~+~${\cal L}_{G_8p^4}^{\Delta S=1}(22)$
~+~${\cal L}_{G_{27}p^4}^{\Delta S=1}(23)$
&   $L=1$ \\[3pt]
~+~${\cal L}_{e^2p^2}^{\rm em}(14)$ ~+
~${\cal L}_{G_8e^2p^2}^{\rm emweak}(15)$~+~
${\cal L}_{e^2p}^{\rm leptons}(4)$ &  \\[10pt] 
~+~${\cal L}_{p^6}^{\rm even}(90)$  & $L=2$ \\[3pt]
%&  \\[3pt] 
\hline
\end{tabular}
$$
\label{tab:EFTSM}
\end{table}

The OPE leads to the standard classification of
$K$ decays: nonleptonic decays correspond to four-quark
operators and semileptonic decays are governed by matrix elements 
of mixed quark-lepton operators. For the actual decays, this
correspondence is however not one-to-one. The standard classification 
applies directly to decays of the type $K \to n \pi$ (nonleptonic; 
${\cal L}_{G_Fp^n}^{\Delta S=1}$) and $K \to n \pi + W^*(\to l \nu_l)$ 
(semileptonic; ${\cal L}_{p^n}$). On the other hand, decays of the 
type $K \to n \pi + l \overline{l}$ (with $l$ either a charged lepton 
or a neutrino) make up an additional class. In this case, both
semileptonic and nonleptonic operators may contribute via their
respective effective Lagrangians. To illustrate this
class of decays, I consider two prominent examples.

For the decays $K \to \pi \nu \overline\nu$, the nonleptonic decay chain
$K \to \pi Z^*(\to \nu \overline\nu)$ induced by  ${\cal L}_{G_Fp^n}^
{\Delta S=1}$ is strongly suppressed: these decays are almost
exclusively given by matrix elements of semileptonic operators
and are therefore classified as short-distance dominated.
The situation is different for the decays $K \to \pi l^+ l^-$ 
where the nonleptonic mechanism $K \to \pi \gamma^*(\to l
\overline{l})$ is important (long-distance dominance). Matrix elements
of the corresponding semileptonic operators are only relevant for CP
violation. 

\begin{center} 
\underline{Current CHPT activities in kaon physics}\footnote{For lack 
of space, references to the original work are omitted here; instead, I 
refer to corresponding contributions to these Proceedings.}
\end{center} 
\begin{itemize} 
\item Systematic higher-order calculations including two-loop
amplitudes for semileptonic decays ($\to$ Sec.~\ref{sec:pipi}).
\item Dispersion theoretic methods for nonleptonic decays ($\to$
Colangelo, Paschos).
\item Methods for determining low-energy constants ($\to$ D'Ambrosio,
de Rafael).
\item CHPT and lattice ($\to$ Golterman).
\item Systematic inclusion of isospin violation and electromagnetic
corrections, both for semileptonic ($\to$ Sec.~\ref{sec:radcorr})
and nonleptonic decays ($\to$ Donoghue, Gardner).
\end{itemize} 

Although I will concentrate in the following on semileptonic decays,
I want to emphasize the importance of isospin violation and
electromagnetic corrections for the dominant (nonleptonic) decays
$K \to 2\pi, 3\pi$. Isospin breaking is
estimated to be of the following size in $K$ decays:
\begin{center}
\begin{tabular}{|c|c|}\hline
%& \\
$O(m_u-m_d)$ & $O(\alpha)$ \\ 
%& \\ 
\hline 
%& \\
$\displaystyle\frac{M_{K^0}^2 - M_{K^+}^2}{M_K^2} \sim 1.5 \%$ & 
$\displaystyle\frac{M_{\pi^+}^2 - M_{\pi^0}^2}{M_K^2} \sim 0.5 \%$ \\
%& \\
\hline  
\end{tabular}
\end{center} 

The effects of $O(m_u-m_d)$ and $O(\alpha)$ are comparable in size and 
they are small.
However, there is a possible strong enhancement in the subdominant
$\Delta I=3/2$ amplitudes because of the $\Delta I = 1/2$ rule, e.g., 
in the $K \to \pi\pi$ amplitude $A_2$:
\begin{equation} 
A_2^{\rm ind}/A_2 \sim \frac{M_{K^0}^2 - 
M_{K^+}^2}{M_K^2} \cdot \frac{A_0}{A_2} \sim 0.35 .
\end{equation} 
Here, the standard ratio $A_0/A_2 \simeq 22$ (assuming isospin
conservation) is used.

In this connection, I want to comment on a recent reanalysis of
$K \to 2 \pi, 3 \pi$ by Cheshkov \cite{chesh} on the basis of the
CHPT amplitudes to $O(p^4)$ \cite{KMW}. Both the (experimental) 
isospin amplitudes and slope parameters and the corresponding CHPT
quantities assume isospin conservation. Therefore, the observed
impressive agreement \cite{chesh} between theory and experiment may be 
somewhat elusive and certainly does not imply that isospin violation 
is negligible. On the
contrary, the neglect of isospin breaking probably hides
interesting physics in the subdominant $\Delta I=3/2$ quantities.

\section{Semileptonic decays: $K_{l4}$ and $\pi\pi$ scattering}
\label{sec:pipi}
\subsection{$\pi\pi$ scattering}
The final state interaction of the two pions in $K_{e4}$ decays
allows for the extraction of the phase shift difference
$\delta_0^0(s)-\delta_1^1(s)$. Recent theoretical work and new
experimental results have led to major improvements.

The CHPT amplitudes to $O(p^6)$ have been known for some 
time \cite{BCEGS}. The more recent dispersion theoretic 
analysis \cite{ACGL} (Roy equations) is a priori completely independent 
of QCD. With high-energy ($\sqrt s \ge 0.8$ GeV) scattering data as
input, it yields $S$ and $P$ waves at low energies in terms of only two
subtraction constants that can be chosen as the $S$-wave scattering
lengths $a_0^0, a_0^2$. The phase shifts ($l \le 1$) and the remaining
threshold parameters are then predicted \cite{ACGL} with amazing
accuracy in terms of $a_0^0, a_0^2$.

\begin{figure}[ht]
  \vspace{6.5cm}
%  \special{psfile=d0md1.eps voffset=-220 hoffset=-30
%    hscale=40 vscale=40 angle=0}
  \includegraphics{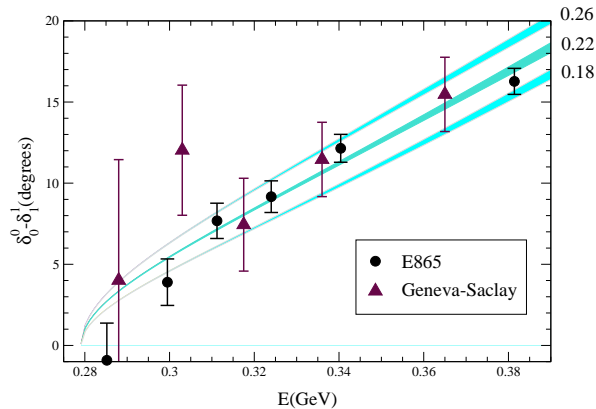}
  \caption{\it Phase shift difference $\delta^0_0 - \delta^1_1$ for
  different values of $a^0_0$ \cite{CGL01b}.}
    \label{d0md1} 
\end{figure}

In the next step, Colangelo et al. \cite{CGL01a} matched the Roy
solutions to the CHPT amplitudes. With some additional
input (most importantly, the LECs $l_3, l_4$ of $O(p^4)$ ), 
the $S$-wave scattering lengths were determined as 
\begin{eqnarray}
\label{a0i}
a_0^0 &=& 0.220 \pm 0.005 \\*
a_0^2 &=& -0.0444 \pm 0.0010 ~,
\end{eqnarray} 
fixing in turn the phase shifts at low energies. Comparing successive
orders, one finds that the chiral expansion of the $\pi\pi$ amplitude
``converges'' well. The weak link in this argument was the standard
CHPT value for $l_3$ that is not beyond discussion. To close this
loophole, $l_3$ was taken as a free parameter
to be determined \cite{CGL01b} from a fit to the new data from
BNL-E865 \cite{BNL}. Including the old data of the Geneva-Saclay
experiment \cite{Rosselet}, the best fit value for $a_0^0$ was found
to be $a_0^0=0.221$, in complete agreement with (\ref{a0i}). The
dependence on $a_0^0$ is shown in Fig.~\ref{d0md1} together with the
available experimental results for the phase shift difference. This 
agreement in turn corroborates the usual value of $l_3$ and  
the standard low-energy expansion scheme corresponding to a large 
quark condensate. At least for chiral $SU(2)$, the
motivation for a generalized scheme \cite{Stern} with a
substantially smaller quark condensate has evaporated.

With the level of accuracy reached (both in CHPT and in the
dispersive analysis), isospin breaking must be included. This is
partly available for $\pi\pi$ scattering \cite{pipiiso} but not yet
for the actual $K_{e4}$ decays.

\subsection{$K_{e4}$ form factors}
Whereas the vector form factor $H$ is dominated by the leading
contribution of $O(p^4)$ due to the chiral anomaly, there are
substantial corrections \cite{ABT01} of $O(p^6)$ to the axial form 
factors $F,G$. The impact of
those corrections can be seen in the updated values \cite{ABT01} of 
some of the LECs of $O(p^4)$ collected in Table \ref{tab:Linew}. 
Although the errors in the last column do not include theoretical 
uncertainties it is clear that the $O(p^6)$ corrections induce 
sizable shifts in some of the LECs ($L_2, L_5, L_8$).

\begin{table}[ht]
  \centering
  \caption{ \it Phenomenological values of $L^r_i(M_\rho)\times
  10^{3}$. The new values in the last column are from the work of
  Amoros et al. \protect\cite{ABT01}.} 
  \vskip 0.1 in
\begin{tabular}{|c||r|r|}  \hline
i & 1995 \hspace*{.37cm} &  ABT 2001 \hspace*{.1cm} \\
\hline
  1  &  0.4 $\pm$ 0.3 & 0.43 $\pm$ 0.12 \\
  2  &  1.35 $\pm$ 0.3  &  0.73 $\pm$ 0.12 \\
  3  &  $-$3.5 $\pm$ 1.1 & $-$2.35 $\pm$ 0.37 \\
  5  &  1.4 $\pm$ 0.5   &  0.97 $\pm$ 0.11 \\
  7  &  $-$0.4 $\pm$ 0.2  &  $-$0.31 $\pm$ 0.14 \\
  8  &  0.9 $\pm$ 0.3  &  0.60 $\pm$ 0.18 \\
\hline
\end{tabular}
\label{tab:Linew}
\end{table}

\section{Isospin violation and electromagnetic corrections in $K_{l3}$
decays}\label{sec:radcorr}
The most precise determination of the CKM matrix element $V_{us}$
comes from $K_{e3}$ decays. The present status \cite{PDG} indicates 
a possible problem (2.2 $\sigma$) for three-generation mixing:
\begin{center} 
\begin{tabular}{|l|c|}
\hline 
& $|V_{us}|$ \\
\hline 
$K_{l3}$  & 0.2196 $\pm$ 0.0023 \\
$|V_{ud}|$ + unitarity & 0.2287 $\pm$ 0.0034 \\
\hline
\end{tabular} 
\end{center} 
Isospin breaking is an essential ingredient \cite{LR84} for 
the extraction of $V_{us}$. However, a complete calculation
to $O(p^4,(m_u-m_d)p^2,e^2p^2)$ has been undertaken only 
recently \cite{CN01}. The relevant Lagrangians of Table \ref{tab:EFTSM}
are ${\cal L}_{p^n}(n\le 4)$ and ${\cal L}_{e^2p}^{\rm
leptons}$ \cite{Knecht}. The calculation of isospin conserving
corrections of $O(p^6)$ is also under way \cite{Bijetal}.

Let me concentrate here on the radiative corrections for $K^+_{l3}$
decays \cite{CN01}. For $\alpha=0$, the decay distribution takes the 
form
\begin{equation} 
A_1^{(0)}(t,u) f_{+} (t)^2  + 
A_2^{(0)}(t,u) f_{+}(t) f_{-}(t)
 + A_3^{(0)}(t,u) f_{-}(t)^2
\label{kl30}
\end{equation}
with kinematical functions $A_i^{(0)}(t,u)$ and form factors $f_\pm(t)$
in terms of the Dalitz variables $t=(p_K-p_\pi)^2$, $u=(p_K-p_l)^2$.

\begin{figure}[ht]
  \vspace{6.0cm}
%  \special{psfile=kl3rc.ps voffset=-220 hoffset=-30
%    hscale=50 vscale=50 angle=0}
  \includegraphics{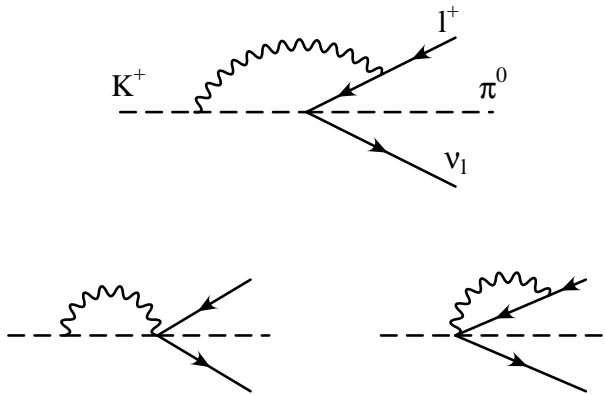}
  \caption{\it
 Radiative corrections for $K^+_{l3}$ decays.}
    \label{kl3rc} 
\end{figure}

Radiative corrections involve the diagrams of Fig.~\ref{kl3rc} and
Bremsstrahlung of soft photons. The final result is a decay
distribution of the same form as (\ref{kl30}) with
\begin{eqnarray} 
f_\pm (t) &  \longrightarrow & f_\pm (t,u)  \\ 
A_i^{(0)} (t,u) &  \longrightarrow & A_i (t,u) ~. 
\end{eqnarray} 
The structure dependent corrections involving the interplay of QCD and
QED are contained in $f_\pm (t,u)$ whereas the universal QED
corrections (Coulomb part of loop corrections + Bremsstrahlung) appear 
in the modified kinematical functions $A_i(t,u)$. The latter depend of
course on the experimental conditions.

Using this representation \cite{CN01}, the form factors $f_\pm (t,u)$
can be extracted from the experimental data in a model independent
way. This will allow for a more reliable determination of the form
factors ($f_-$ is still poorly known) and also, as a
consequence, of the CKM matrix element $V_{us}$. Numerical
results will be available \cite{CN01} when these Proceedings appear
in print. 

\section{Conclusions}
The combination of OPE and CHPT provides a comprehensive framework for
the analysis of {\bf all} $K$ decays:
\begin{itemize} 
\item
The structure and the renormalization of ${\cal L}_{\rm eff}^{\rm SM}$ 
are well understood including electromagnetic corrections.
\item
Significant advances have been achieved in the analysis of
$\pi\pi$ scattering (related to $K_{e4}$ decays).
\item
$O(p^6)$, isospin breaking and electromagnetic corrections are being 
completed for semileptonic decays.
\item
More work is needed to determine many of the LECs in order to make 
the scheme even more predictive.
\end{itemize}

\section{Acknowledgements}
I thank Vincenzo Cirigliano and Helmut Neufeld for 
information about their unpublished work \cite{CN01}.

\end{document}